# Resonance Beyond Frequency-Matching

Zhenyu Wang (王振宇)[1], Mingzhe Li (李明哲)[1,2], & Ruifang Wang (王瑞方)[1,2*]

[1] Department of Physics, Xiamen University, Xiamen 361005, China.

[2] Institute of Theoretical Physics and Astrophysics, Xiamen University, Xiamen 361005, China.

*Corresponding author.   wangrf@xmu.edu.cn

Resonance, defined as the oscillation of a system when the temporal frequency of an external stimulus matches a natural frequency of the system, is important in both fundamental physics and applied disciplines. However, the spatial character of oscillation is not considered in the definition of resonance. In this work, we reveal the creation of spatial resonance when the stimulus matches the space pattern of a normal mode in an oscillating system. The complete resonance, which we call multidimensional resonance, is a combination of both the spatial and the conventionally defined (temporal) resonance and can be several orders of magnitude stronger than the temporal resonance alone. We further elucidate that the spin wave produced by multidimensional resonance drives considerably faster reversal of the vortex core in a magnetic nanodisk. Our findings provide insight into the nature of wave dynamics and open the door to novel applications.

# I. INTRODUCTION

Resonance is a universal property of oscillation in both classical and quantum physics[1,2]. Resonance occurs at a wide range of scales, from subatomic particles[2,3] to astronomical objects[4]. A thorough understanding of resonance is therefore crucial for both fundamental research[4-8] and numerous related applications[9-12]. The simplest resonance system is composed of one oscillating element, for instance, a pendulum. Such a simple system features a single inherent resonance frequency. More commonly, a resonator contains interacting elements and has multiple resonance frequencies. Each resonance frequency corresponds to a normal mode[1] that is characterized by unique spatial variation of the oscillation amplitude and of the phase that are determined by the resonator's geometry and boundary conditions. There have been extensive studies on the tunability of resonance frequency and the accessibility of local modes by geometrical means in artificial nanostructures[6,7,12,13], while the spatial feature of the normal modes receives little attention.

To date, in the conventional definition of resonance, the only criterion is whether the temporal frequency of the external stimulus is equal to one of the resonance frequencies of the system, and the spatial character of the normal modes is ignored. Since frequency describes the periodic pattern of oscillation in the time domain, we choose to use a more specific term "temporal resonance" for such defined phenomenon. In this work, we reveal the generation of spatial resonance, i.e. resonance in the space domain, other than the temporal resonance, when we align the spatial distribution of the external stimulus with the space pattern of a normal mode. The complete resonance, which we call multidimensional resonance, must incorporate both temporal and spatial resonance. The temporal resonance has low capability of stimulating all modes but the fundamental mode, as a result of neglecting the spatial character of normal modes. In contrast, multidimensional resonance is efficient for all the normal modes; thus it can be several orders of magnitude stronger than the temporal resonance alone.

We conduct micromagnetic simulations (see Appendix A) and analytical derivations on the behaviour of spin waves in a ferromagnetic nanodisk and elastic waves in a two-dimensional circular membrane, respectively. The interplay between spin waves and magnetization reversal in small magnetic structures is of particular interest to both fundamental physics and applications in high-speed data storage devices[14-16] and logic circuits[17-20]. In ferromagnetic nanodisks a unique spiral spin configuration called vortex state[21-23] is favorable due to the competition between magnetostatic and exchange interactions. The magnetization at the vortex core can point either upwards (polarity $p = +1$) or downwards ($p = -1$); therefore, a magnetic vortex can be used as a data storage element to carry one bit of information. Three classes of spin wave modes, namely, gyrotropic[24], azimuthal and radial[25-27], have been discovered in magnetic disks. We focus on the radial spin wave mode of the ferromagnetic disk because it is analogous to the radially symmetrical mode[28] in a two-dimensional elastic membrane, thus facilitating direct comparison of the calculation results in the two systems. Recent studies have shown that the resonant radial mode is capable of reversing the core polarity through a novel mechanism[29-32], wherein the core switching is attributable to the breathing nature of spin wave, instead of the well-known vortex-antivortex annihilation process found in the gyrotropic- and azimuthal-mode-driven core reversal[14,33-38].

Our calculations demonstrate spatial resonance in both the magnetic and mechanical resonators under the space-pattern-matching condition. This suggests that spatial resonance does not depend on material and, similar to the temporal resonance, is a universal property of oscillation systems. We also show that multidimensional resonance, i.e. complete resonance in both time and space domain, drives a markedly stronger radial spin wave than does temporal resonance alone and increases the core reversal speed by more than 500%. Additional results for mechanical and magnetic strings are presented in Appendix B and C, respectively, as demonstrations of the multidimensional resonance in spatially one-dimensional systems.

## II. NORMAL MODES AND SPACE PATTERN

Consider the forced small amplitude vibrations of a membrane stretched in a rigid circular frame; the equation of motion is given by[28,39]

$$\frac{\partial^2 W}{\partial t^2} - a^2 \nabla^2 W = \frac{p(\rho,\varphi,t)}{\tilde{\rho}} = g(\rho)\cos m\varphi \sin kat. \tag{1}$$

Here, $\nabla^2$ is the Laplace operator, $W(\rho,\varphi,t)$ is the vertical displacement of a membrane with radius $\rho_0$ and mass per unit area $\tilde{\rho}$, $a=\sqrt{T/\tilde{\rho}}$ is the propagation velocity of the transverse waves, and $T$ is the isotropic tension in the membrane with dimensions of force per unit length. To facilitate discussing the efficiency of different driving forces (both spatially and temporally), the external force acting normal to the membrane is assumed to have surface density $p(\rho,\varphi,t)$ taking the form of the right side of equation (1) and subject to normalization condition $\int_0^{2\pi} \int_0^{\rho_0} |g(\rho)\cos m\varphi| \rho d\rho d\varphi = c\pi\rho_0^2$, where $c$ is a constant.

For a stretched membrane initially at rest, the initial conditions are $W(\rho,\varphi,0) = 0$ and $[\partial W/\partial t]_{t=0} = 0$, whereas the boundary conditions are $W(\rho_0,\varphi,t) = 0$ and $[\partial W/\partial t]_{\rho=\rho_0} = 0$. The solution to equation (1) satisfying these initial-boundary value conditions is obtained by separation of variables, with the result

$$W(\rho,\varphi,t) = \frac{\cos m\varphi}{a} \sum_{n=1}^{\infty} \frac{J_m(k_n^{(m)}\rho)}{k_n^{(m)}} I_n L_n(t), \qquad (2)$$

in which

$$I_n = \frac{2}{\rho_0^2 [J_{m+1}(k_n^{(m)}\rho_0)]^2} \int_0^{\rho_0} g(\rho) J_m(k_n^{(m)}\rho) \rho d\rho \qquad (3)$$

and

$$L_n(t) = \int_0^t \sin ka\tau \sin k_n^{(m)} a(t-\tau) d\tau. \qquad (4)$$

Here, $J_m(x)$ is the Bessel function of the first kind of order $m$, $k_n^{(m)} = x_n^{(m)}/\rho_0$, and $x_n^{(m)}$ is the $n$-th non-negative root of $J_m(x)$. Expression (2) shows that the stimulated vibration is a superposition of the normal modes $J_m(k_n^{(m)}\rho)\cos m\varphi$, as indicated by spatial pattern index $(n,m)$, where $n = 1, 2, ..., \infty$ and $m = 0, 1, ..., \infty$. $J_m(x)$ changes in sign whenever $x$ moves across a node at $x_n^{(m)}$, resulting in phase reversal. The angular factor of the normal mode, $\cos m\varphi$, has opposite phases on either side of the nodal line at $\varphi = \pm\pi/2m, \pm 3\pi/2m, ..., \pm(2m-1)\pi/2m$. Thus, space can be partitioned into phase zones by nodal lines, forming a space pattern that is unique to the index $(n,m)$ (Figure 1(a)).

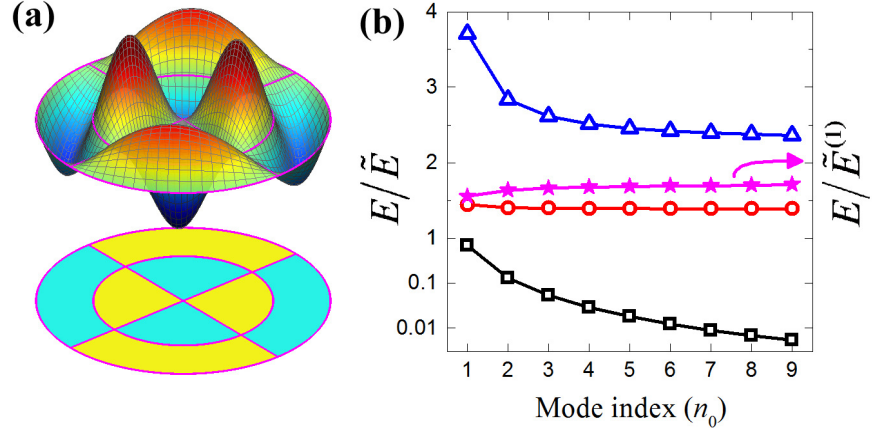

FIG. 1. Illustration of a radially asymmetrical mode $(n, m) = (2, 2)$ of an elastic membrane with a fixed boundary (a), and membrane energy growth versus mode index for different stimuli (b). (a) The mode shape is described by $J_2(k_2^{(2)}\rho)\cos 2\varphi$. There are two circular nodal lines along the azimuthal direction and two straight nodal lines along the radial direction. The mode changes in sign across a nodal line. All nodal lines are shown in pink. The plane view of the mode shows that the sample is partitioned by nodal lines into eight phase zones. The yellow zones are in the opposite oscillation phase to the cyan zones. (b) The black square, red circle and blue triangle lines show the membrane energy growth expressed as $E/\tilde{E}$ versus mode index for different stimuli in equations (6), (7) and (10), respectively. The pink star line illustrates the membrane energy growth expressed as $E/\tilde{E}^{(1)}$ versus mode index for the stimulus in equations (11).

Now, we turn our attention to the micromagnetic simulation results on the radial spin wave mode of a 300 nm diameter and 5 nm thick permalloy ($Ni_{80}Fe_{20}$) nanodisk. In the static state, the disk forms a vortex state (Fig. 2(a)) with core polarity equal to +1. To study the radial mode, we apply a sinc function field over the entire disk to excite the spin wave. The resulting temporal oscillation of the z-component magnetization averaged over the whole disk, $<m_z>$, is given in Fig. 2(b). Next, we perform fast Fourier transform (FFT) to obtain the amplitude spectrum in the frequency domain[27,40,41]. Seven primary peaks corresponding to radial modes ($n$ = 1, 2, …, 7) appear at resonant frequencies of $f_n$ = 6.8, 9.8, 12.6, 15.9, 19.6, 24.0 and 29.0 GHz (Fig. 2(c)). In the FFT-amplitude spatial distribution diagrams for the first four normal modes (Fig. 2(d)), we see clear quantization of the spin wave, and the number of nodes along radial direction corresponds to the mode index $n$. Additionally, when plotting phase diagrams for these modes, we observe high spatial uniformity at $n$ = 1 but phase discontinuity of $\pi$ across nodal lines for $n$ > 1, as expected for the standing wave nature of the normal modes. Therefore, the entire sample can be divided into ring-shaped phase zones similar to the case of the membrane.

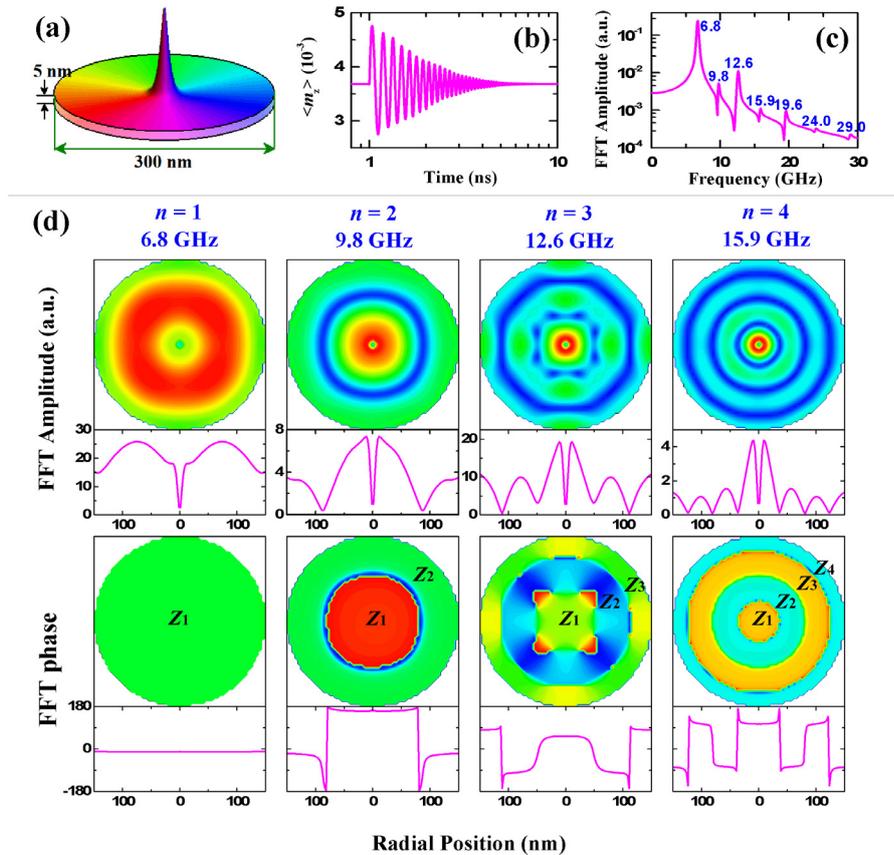

FIG. 2. A magnetic nanodisk and its normal modes. (a) A nanodisk of indicated dimensions

and an initial vortex ground state in which the vortex-core magnetization is upwards. (b) Average $m_z$ component, $<m_z>$, over the entire disk versus time after applying the out-of-plane sinc function field that is expressed as $H_{\text{sinc}} = A\sin[2\pi f(t-t_0)]/2\pi f(t-t_0)$ with $A = 50$ Oe, $f = 100$ GHz and $t_0 = 1$ ns. (c) FFT amplitude spectrum of the sample with seven identified resonance frequencies. (d) Plane view images of the spatial distribution FFT amplitude (upper panels) and phase (lower panels) for the first four radial modes, along with profiles across the center position. The sample is divided into ring-shaped phase zones as indicated by the labels $Z_1$, $Z_2$, ….

## III. SPATIAL RESONANCE AND MULTIDIMENSIONAL RESONANCE

We first theoretically establish the spatial resonance in the elastic membrane. Because expression (3) means the projection of $g(\rho)$ on each mode shape $J_m(k_n^{(m)}\rho)$ and because $g(\rho)$ denotes the radial distribution of the driving force, $I_n$ implies space-domain resonance in the radial dimension, when $g(\rho)$ matches the space pattern of $J_m(k_n^{(m)}\rho)$. Similarly, expression (4) for $L_n(t)$ is the convolution of the temporal variations of the driving force and the generated wavelet (i.e., a response to stimulation in the time domain) and implies time-domain resonance. A higher degree of similarity between the response and stimulation as functions of time gives rise to a larger convolution $L_n(t)$, which is given by

$$L_n(t) = \begin{cases} \dfrac{k\sin k_n^{(m)}at - k_n^{(m)}\sin kat}{a(k+k_n^{(m)})(k-k_n^{(m)})} & \text{if } k \neq k_n^{(m)} \\ -\dfrac{t\cos k_n^{(m)}at}{2} + \dfrac{\sin k_n^{(m)}at}{2k_n^{(m)}a} & \text{if } k = k_n^{(m)} \end{cases} ; \quad (5)$$

thus, only temporal resonance causes a time accumulation effect.

For a spatially uniform stimulus with conventional, temporal, resonance, i.e., $m=0$ and $p(\rho,\varphi,t)/\tilde{\rho} = c\sin k_{n_0}^{(0)}at$, we have $I_n = 2c/\left[x_n^{(0)}J_1\left(x_n^{(0)}\right)\right]$. Then, the energy in the membrane grows as

$$E = \int_0^{2\pi}\int_0^{\rho_0} \tilde{\rho}(\partial W/\partial t)^2 \rho d\rho d\varphi = \tilde{E}\left[2/x_{n_0}^{(0)}\right]^2 + HFOTs, \tag{6}$$

where $\tilde{E} = \left(\pi\rho_0^2\tilde{\rho}c^2/8\right)t^2\left(1-\cos 2k_{n_0}^{(0)}at\right)$, and the high frequency oscillating terms ($HFOTs$) determine fine structure of the energy curve with no influence on its general trend. Figure 1(b) shows that the energy growth rate quickly drops for larger $n_0$ as $E/\tilde{E} \approx 4/\left[\pi(n_0-1/4)\right]^2$ because the stimulus, which is uniform in space and has the spatial pattern index $(1,0)$, is off-resonance with space pattern $(n_0,0)$ when $n_0 \neq 1$; and is farther from the resonance for larger $n_0$.

For a radially symmetrical stimulus taking the spatial resonance into consideration, for example, $m=0$ and $p(\rho,\varphi,t)/\tilde{\rho} = \left(c/s_{n_0}\right)J_0\left(k_{n_0}^{(0)}\rho\right)\sin k_{n_0}^{(0)}at$, the normalization condition demands $s_{n_0} = 2\left[x_{n_0}^{(0)}\right]^{-2}\sum_{i=1}^{n_0}(-1)^{i-1}h_i$, where $h_i = x_i^{(0)}J_1\left(x_i^{(0)}\right) - x_{i-1}^{(0)}J_1\left(x_{i-1}^{(0)}\right)$. Then, $I_n$ equals $c/s_{n_0}$ if $n=n_0$ and $0$ if $n\neq n_0$, which shows that a stimulus tends to generate a wave similar to itself and that the normal modes are mutually orthogonal. Hence, the energy in the membrane grows as

$$E = \tilde{E}\left[J_1\left(x_{n_0}^{(0)}\right)/s_{n_0}\right]^2 + HFOTs. \tag{7}$$

Figure 1(b) shows that the energy grows at similar rates for different $n_0$ (limiting at $E = 1.388\tilde{E}$ for large $n_0$). In this case, the stimulus has spatial pattern index $(n_0,0)$ and is on-resonance with space pattern $(n_0,0)$. The energy growth expressed by equation (7) is at least one order of magnitude larger than that expressed by equation (6), since the latter case is off-resonance for $n_0 > 1$.

It is technically difficult to set the driving force varying as a Bessel function in the radial dimension. For ease of implementation, we employ a delta-function-like stimulus as follows

$$\frac{p(\rho,\varphi,t)}{\tilde{\rho}} = \frac{c\rho_0^2}{s_{n_0}\left[x_{n_0}^{(0)}\right]^2} \sum_{i=1}^{n_0} h_i \frac{\delta(\rho-\rho_i)}{\rho} \cos m\varphi \sin k_{n_0}^{(m)} at, \qquad (8)$$

where $\rho_i = \rho_0\left(X_i^{(m)}/x_{n_0}^{(m)}\right)$, $X_i^{(m)}$ is the $i$-th extremum of $J_m(x)$, and the coefficient comes from the normalization condition. The radially symmetrical case of this stimulus has the same space pattern $(n_0, 0)$ as the preceding example, except that the force is concentrated on each mode extremum and is expected to be more effective. Now we have

$$I_n = \frac{2c}{s_{n_0}\left[x_{n_0}^{(0)}\right]^2} \times \begin{cases} \sum_{i=1}^{n_0} h_i J_0\left(X_i^{(0)}\right) & \text{if } n_0 = n, \\ \sum_{i=1}^{n_0} h_i J_0\left(\frac{x_n^{(0)}}{x_{n_0}^{(0)}} X_i^{(0)}\right) & \text{if } n_0 \neq n, \end{cases} \qquad (9)$$

and

$$E = \tilde{E}\left[\frac{\sum_{i=1}^{n_0} h_i J_0\left(X_i^{(0)}\right)}{J_1\left(x_{n_0}^{(0)}\right)\sum_{i=1}^{n_0}(-1)^{i-1} h_i}\right]^2 + HFOTs. \qquad (10)$$

Referring to Fig. 1(b), the stimulus with $n_0 = 1$ in this case (i.e., a point stimulus concentrated on the membrane center) is the most effective. And the energy growth rate is enhanced by 62% to 157% for $1 \leq n_0 \leq 9$, comparing to the case expressed by equation (7).

In the case of a magnetic nanodisk, we choose the $n = 2$ radial mode to demonstrate the spatial resonance of the magnetic vortex by numerical calculations. We apply a simple harmonic field with the frequency tuned to the second modal frequency, i.e., $f = f_2$, and a small oscillation amplitude of 10 Oe for 10 ns. The spatial distribution of the field phase aligns (partially or fully) with the $n = 2$ mode. The FFT amplitude and phase images for the $n = 2$ mode after application of the resonant frequency field is shown in Fig. 3. The nodal line at $\rho = 87.5\ nm$ partitions the sample into two phase zones: $Z_1$ for $\rho \leq 87.5\ nm$ and $Z_2$ for $\rho > 87.5\ nm$. Figures 3(a) and 3(b) show the FFT amplitude and phase diagrams when the field is localized in $Z_1$ and $Z_2$, respectively. The FFT amplitude has a slightly higher value in Fig. 3(b), which can be explained by the larger area of $Z_2$ than $Z_1$. The FFT phase variations along $\rho$ are opposite. As the external field reverses its oscillation phase, the resultant FFT phase images are reversed, whereas the amplitude profile remains intact (Fig. 3(c)). The same effect exists for an elastic membrane. Refer to expressions (2) and (9); the factor $I_n$ in the vibration $W(\rho, \varphi, t)$ changes in sign whenever the stimulus moves across a node, resulting in a phase reverse. We can thus interpret the small oscillation amplitude generated by a uniform global field (which produces temporal resonance only) in Fig. 3(d) as the result of the destructive addition of oscillations in Figs. 3a and 3b. In sharp contrast, we observe a much larger oscillation amplitude in Fig. 3(e) when the field distribution matches the space pattern of the mode, thus satisfying the spatial resonance condition, and the resultant oscillation corresponds to the constructive addition of oscillations in Figs. 3a and 3c. This reasoning is also applicable to explaining why the multidimensional resonance is far more energetic than the temporal resonance for the membrane, as shown in Fig. 1(b).

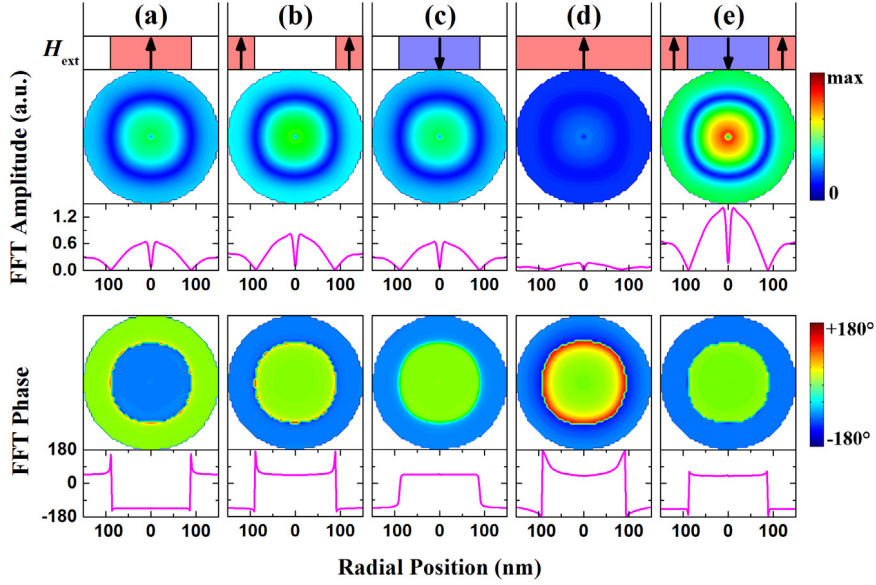

FIG. 3. FFT amplitude and phase images, along with profiles across the center position, for mode $n = 2$ of the nanodisk under different external fields. The simple harmonic stimulus is perpendicular to the disk surface and has the form $H = a\sin(2\pi f_2 t + \varphi_0)$, where $a$ is the oscillation amplitude of 10 Oe, $f_2$ is the second modal frequency of 9.8 GHz, and $\varphi_0$ is the initial phase. Arrows in the top panel schematically show the initial phase of the external field. Upward arrows denote $\varphi_0 = 0$, and downward arrows denote $\varphi_0 = \pi$. (a) The field is localized in the $Z_1$ phase zone. (b) The field is localized in $Z_2$. The FFT phase image is opposite to the one in (a). (c) The external field is confined to $Z_1$ but with the opposite initial phase to the phase in (a). The resultant FFT phase image is very similar to (b). (d) The field is uniformly applied over the entire disk. The resultant FFT amplitude is small. (e) The field distribution matches the space pattern of mode $n = 2$ and satisfies the multidimensional resonance condition. As a result, the FFT amplitude is much larger than in (d).

Figure 4 shows the growth in total energy ($\Delta E$) of the seven radial modes of our magnetic nanodisk under temporal and multidimensional resonance conditions. The external field oscillation amplitude is small (10 Oe) in all cases to suppress nonlinear dynamics. To make the results comparable, the summation of the magnetic flux (absolute value) in all phase zones is the same in the micromagnetic simulations. For the temporal resonance in Fig. 4(a), we observe that the fundamental mode ($n = 1$) accumulates energy at a rate of more than one order higher than the other modes. The general trend is that for odd $n$, the rate dramatically decreases as $n$ increases, similar to the temporal resonance of the membrane shown in Fig. 1(b). For even $n$, the corresponding energy is close to zero. However, multidimensional resonance produces strikingly different results in Fig. 4(b). Here, the external field fits the space pattern of the corresponding radial mode, but the field amplitude is uniformly distributed over the entire disk. The sample increases in energy at a faster rate for higher modes, so the fundamental mode has the lowest efficiency. The system gains 3.5 times more energy for mode $n = 7$ than $n = 1$. Furthermore, modes $n = 2$ to 7 increase in energy much faster than their temporal resonance counterparts shown in Fig. 4(a). The ratio of the two energy gain values is a staggering 1856 for $n = 7$. The fundamental mode has no difference in energy growth because it is the only mode with a globally uniform phase. The system receives an additional energy boost when the spatial variation of the external field matches both the space pattern and the mode amplitude profile, as shown in Fig. 4(c). The energy growth rates are 4% to 103% higher than those in Fig. 4(b).

Finally, we observe the highest rate (19 to 34 percent higher for all modes than their counterparts in Fig. 4(c)) when we concentrate the field at the antinodes (Fig. 4(d)) [42].

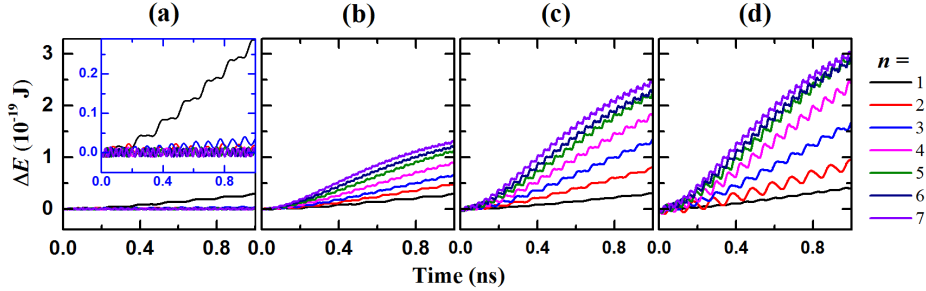

FIG. 4. Time evolution of energy growth for the radial modes of the magnetic nanodisk. The sample is in temporal resonance in (a). The inset displays energy gains at a reduced scale. In (b), (c) and (d), the sample satisfies both temporal and spatial resonance conditions. In addition, the spatial distribution of the external field is uniform in (b), matches the mode amplitude profile in (c) and is concentrated on the antinodes in (d).

## IV. ULTRAFAST REVERSAL OF THE MAGNETIC VORTEX CORE

Because multidimensional resonance excites a strikingly stronger spin wave than does temporal resonance, we expect multidimensional resonance to facilitate a much faster vortex core reversal. Figure 5 displays the core reversal time for the first seven radial modes when the spatial distribution of the external field matches the space pattern of the corresponding mode and the field amplitude is kept uniform at 300 Oe within all phase zones. The reversal time $t_{sw}$ is 588 ps for $n = 1$, decreases dramatically to 288 ps for $n = 2$, and continues to decrease at slower pace for higher modes until reaching 93 ps for $n = 7$ (Fig. 6), which marks a 532% faster reversal speed than for the fundamental mode. These results are in sharp contrast to references [30] and [32], where the authors produced temporal resonance only and the reversal time increases quickly for higher modes. In addition to achieving sub-100 picosecond reversal, multidimensional resonance greatly lowers the threshold field for core reversal. Figure 5 shows that the threshold $H_{thr}$ is 285 Oe for $n = 1$ and steadily decreases to 55 Oe at $n = 5$. The unexpected, slightly higher values at $n = 6$ and 7 are attributable to the reduced accuracy of the numerical calculation for higher modes, in which the ring-shaped phase zones appear more zigzag-like in the micromagnetic simulation.

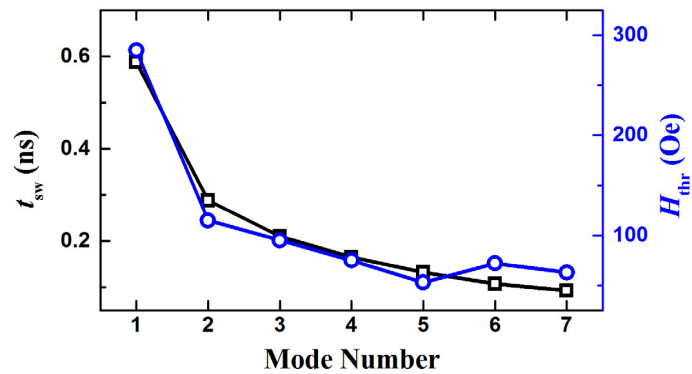

FIG. 5. Variation of the core reversal time $t_{sw}$ and threshold field $H_{thr}$ versus the radial mode index. When studying the reversal time, the field amplitude is uniformly distributed at 300 Oe in all phase zones.

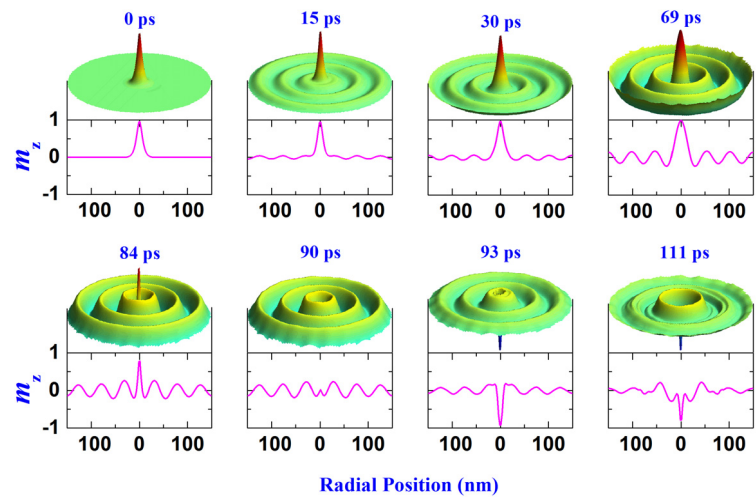

FIG. 6. Snapshot images of the temporal evolution of the $n = 7$ radial mode with the corresponding $m_z$ profiles. The external field satisfies both the temporal and spatial resonance conditions, with the field amplitude uniformly distributed in all seven phase zones at 300 Oe.

## V. DISCUSSION AND OUTLOOK

The circular symmetry of the membrane (and the ferromagnetic nanodisk) is responsible for the non-periodic space pattern in the radial dimension. But for a rectangular membrane[28] or the strings (see Appendix B and C), the geometrical symmetry results in a periodic space pattern for the oscillation modes. Similarly, frequency, i.e. the temporal pattern of oscillation, can also be tuned by confinement of wave in the time domain, or even the geometry of spacetime[43]. Pure periodic oscillation exists only when the oscillation exists for infinitely long time within a flat spacetime. Therefore, the essence of both temporal and spatial resonance is the pattern-matching. Additionally, concentration of stimulus at the mode extremes can considerably enhance the resonance effect, as shown in Fig. 1(b) and Fig. 4.

In the above study on the membrane, for simplicity, we postulate that the stimulus is on-resonance in the angular dimension and discuss only the radial dimension. The factor $\cos m\varphi$ in (1) and (2) implies space-domain resonance in the angular dimension; thus, each space dimension should be given an index, and together these indices constitute the spatial pattern index, which completely describes the spatial character.

For a radially asymmetrical stimulus with spatial resonance, for example, $p(\rho,\varphi,t)/\tilde{\rho} = (c/s_{n_0}^{(m)}) J_m(k_{n_0}^{(m)}\rho) \cos m\varphi \sin k_{n_0}^{(m)} at$ , the normalization condition demands $s_{n_0}^{(m)} = \frac{4}{\pi \rho_0^2} \int_0^{\rho_0} \left| J_m \left( k_{n_0}^{(m)} \rho \right) \right| \rho d\rho$. Hence, the energy in the membrane grows as

$$E = \tilde{E}^{(m)} \left[ J_{m+1} \left( x_{n_0}^{(m)} \right) \middle/ \sqrt{2} s_{n_0}^{(m)} \right]^2 + HFOTs, \tag{11}$$

where $\tilde{E}^{(m)} = \left( \pi \rho_0^2 \tilde{\rho} c^2 / 8 \right) t^2 \left( 1 - \cos 2 k_{n_0}^{(m)} at \right)$. In this case, the stimulus has spatial pattern index $(n_0, m)$ and is on-resonance with space pattern $(n_0, m)$. For $m = 1$, the energy grows at similar rates for different $n_0$ (limiting at $E = 1.712 \tilde{E}^{(1)}$ for large $n_0$) but is distinct from the radially symmetrical case in that the coefficient $E/\tilde{E}^{(1)}$ increases with $n_0$ (Fig. 1(b)). In passing, it is interesting to notice one feature of a radially asymmetrical mode, namely that the membrane center remains still at all times.

In the analytical theory of thin magnetic nanodisks[25], the Bessel function $J_1 \left( k_n^{(1)} \rho \right)$ is the approximate wave function of the radial mode. One key difference between $J_1 \left( k_n^{(1)} \rho \right)$ and the micromagnetic simulation result is that $J_1 \left( k_n^{(1)} \rho \right)$ is zero but the numerical solution is significant at the boundary. The small deviation of the mode shape in the analytical and numerical solutions produces a large difference in spatial resonance. The analytical solution predicts only a slightly higher energy growth rate for larger $n_0$, as given in Fig. 1(b). For comparison, the numerical solution presents a much more energetic spatial resonance for larger $n$, as shown in Fig. 4(c). These results highlight the sensitivity of the spatial resonance to the mode shape, which can be tuned by factors such as sample geometry and composition and boundary conditions.

In summary, our calculations on elastic membranes and magnetic nano-discs demonstrate multidimensional resonance wherein the external stimulus aligns with not only the temporal frequency but also the space pattern of a normal mode. For a spatially uniform stimulus, the external stimuli acting on zones of opposite phases are destructive.

Therefore, the conventional (temporal) resonance is inefficient at exciting all modes but the fundamental mode. In contrast, multidimensional resonance is efficient for all modes and creates a much stronger oscillation because the external stimulus reverses its direction in adjoining zones; thus, all stimuli are constructive. Because the manifestation of wave interference does not depend on matter, multidimensional resonance is expected to be a universal property of oscillation systems. Our research reshapes the theory on resonance and opens a new arena wherein the spatial character of oscillation modes plays a key role. Future studies on ways of tailoring the mode shape may lead to novel technological innovations that exploit the nature of multidimensional resonance.

## ACKNOWLEDGEMENTS

We acknowledge the financial support from the National Natural Science Foundation of China under Grant Nos. 10974163 and 11174238. R.W. is grateful for the useful conversations with K. Shih.

Z. W. and M. L. contributed equally to this work.

## APPENDIX A. METHODS

Our micromagnetic simulations are conducted using LLG Micromagnetics Simulator code[44], which numerically solves the Landau-Lifshitz-Gilbert equation for the dynamic magnetization process. The magnetic nanodisk and nanowire in the model are composed of permalloy ($Ni_{80}Fe_{20}$). We use the typical material parameters for permalloy: saturation magnetization $M_s = 8.0 \times 10^5 \, A/m$, exchange stiffness constant $A_{ex} = 1.3 \times 10^{-11} \, J/m$, Gilbert damping constant $\alpha = 0.01$, and zero magnetocrystalline anisotropy. In the simulations, the mesh cell size is $2.5 \times 2.5 \times 5.0 \, nm^3$ for the nanodisk and $2.5 \times 2.5 \times 2.5 \, nm^3$ for the nanowire.

The total energy of our models includes contributions from the Zeeman, exchange and demagnetization energy.

## APPENDIX B. MULTIDIMENSIONAL RESONANCE IN AN ELASTIC STRING

The forced small amplitude vibration of a stretched finite string is the solution of partial differential equation

$$\frac{\partial^2 W}{\partial t^2} - a_0^2 \frac{\partial^2 W}{\partial x^2} = \frac{f^*(x,t)}{\rho^*} = g(x) \sin k a_0 t, \tag{A1}$$

under initial conditions $W(x,0) = 0$ and $[\partial W / \partial t]_{t=0} = 0$ and boundary conditions $W(x_0, t) = 0$ and $[\partial W / \partial t]_{x=x_0} = 0$, where $x_0$ is the length of the string, $\rho^*$ is its mass per unit length, $a_0 = \sqrt{T_0 / \rho^*}$, $T_0$ is the tension in the string, and $f^*(x,t)$ denotes the linear density of the external driving force which satisfies normalization condition $\int_0^{x_0} |g(x)| dx = c x_0$. The solution is well-known:

$$W(x,t) = \frac{2}{a_0 x_0} \sum_{n=1}^{\infty} \frac{\sin(n \pi x / x_0)}{n \pi / x_0} \int_0^{x_0} g(\xi) \sin \frac{n \pi}{x_0} \xi \, d\xi \int_0^t \sin k a \tau \sin \frac{n \pi}{x_0} a_0 (t-\tau) d\tau.$$
(A2)

For a stimulus with temporal resonance that is uniform in space, i.e., $f^*(x,t)/\rho^* = (c/x_0) \sin(n_0 \pi a_0 t / x_0)$, the energy in the string grows as $E = E^* / n_0^2 + HFOTs$ for odd $n_0$ and is null for even $n_0$, where

$E^* = \left( \rho^* c^2 / \pi^2 x_0 \right) t^2 \left( 1 - \cos 2n_0 \pi a_0 t / x_0 \right)$. The energy, whereas similar to the two-dimensional-space case in growing proportionally to $1/n_0^2$ for odd $n_0$, does not grow with time for even $n_0$. Only when $n_0 = 1$ does spatial resonance condition satisfy; otherwise, the external force acting on adjoining zones are destructive and cancel each other exactly for even $n_0$, or remain one zone for odd $n_0$ (i.e., $\int_0^{x_0} g(x) \sin n_0 \pi x / x_0 \, dx = 0$ for even $n_0$ if $g(x)$ is a constant).

For stimulus with spatial and temporal resonance but uniform in each space zone, i.e.,

$$\frac{f^*(x,t)}{\rho^*} = \frac{4c}{\pi} \sum_{k=1}^{\infty} \frac{1}{2k-1} \sin \frac{n_0 \pi (2k-1) x}{x_0} \sin \frac{n_0 \pi a_0 t}{x_0} \quad (A3)$$

energy in the string grows as $E = E^* + HFOTs$. If the preceding stimulus of complete resonance goes a step further from spatial pattern-fitting to profile-fitting, such that $f^*(x,t)/\rho^* = (c\pi/2) \sin(n_0 \pi x / x_0) \sin(n_0 \pi a_0 t / x_0)$, then the energy grows as $E = \pi^4 E^* / 64$. Finally, calculations show that if the stimulus in each space zone is concentrated on each mode extremum, the energy grows as $E = \pi^2 E^* / 4$. The foregoing three cases are complete resonance; their energy growth does not depend on $n_0$. And obviously, the string receives additional energy boost when greater proportion of the force acts on the mass points with larger oscillation displacement.

## APPENDIX C. MULTIDIMENSIONAL RESONANCE IN A MAGNETIC NANOWIRE

The magnetic nanowire in the model is 300 nm long along *x*, 2.5 nm wide along *y* and 2.5 nm thick along *z*. In the initial state, the magnetization is along +*x* because of the large shape anisotropy of the wire. In addition, we apply a pinning field ($H_{\text{pin}}$) of the order of 1.5 kOe along +x at the two ends to achieve a fixed boundary condition, as shown in the inset of Fig. 7.

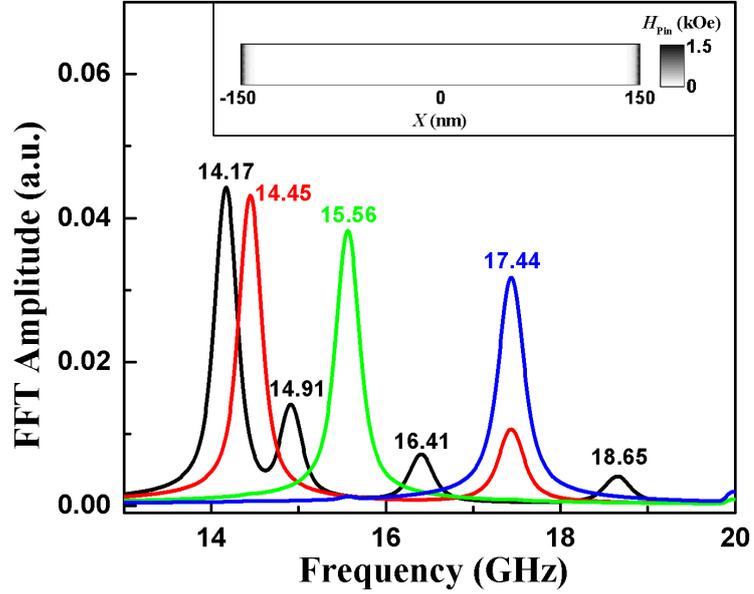

FIG. 7. FFT amplitude spectra of the permalloy nanowire. A globally uniform sinc field produces the spectrum plotted in black, which shows four resonance peaks corresponding to the normal mode, at $n$ = 1, 3, 5, and 7. Even-index modes are excited by the spatial resonance field, i.e., the sinc field, with a spatial distribution that aligns with the space pattern of the mode under study. The FFT spectra plotted in red, green and blue are obtained from the spatial resonance field for $n$ = 2, 4 and 6, respectively. The inset shows the permalloy nanowire with the indicated dimensions and distribution of the pinning field.

To study the normal mode, we apply a sinc function field along $z$ in the form of $H_{\text{sinc}} = A_0 \sin[2\pi f(t-t_0)]/2\pi f(t-t_0)$, with $A_0$ = 50 Oe, $f$ = 100 GHz and $t_0$ = 1 ns, over the entire wire to stimulate a spin wave. Fast Fourier transform (FFT) analysis on the subsequent temporal oscillation of the averaged $m_z$ over the whole wire yields the resultant FFT amplitude in the frequency domain, as shown in Fig. 7. The four peaks at frequencies $f_n$ = 14.17, 14.91, 16.41 and 18.65 GHz correspond to normal modes with indices of $n$ = 1, 3, 5, and 7, respectively. Because the globally uniform field is unable to excite even-index modes, we use a spatial resonance field to selectively drive the individual modes. For instance, to excite the $n = 2$ mode, the external field is in the form of $H_{\text{sinc}}$ within [-150 nm, 0 nm] and $-H_{\text{sinc}}$ within [0 nm, 150 nm]. The resultant FFT amplitude spectra for $n$ = 2, 4, and 6 obtained by this method are given in Fig. 7, which shows the primary resonance peaks at $f_2$ = 14.45 GHz, $f_4$ = 15.56 GHz and $f_6$ = 17.44 GHz, respectively. The spatial resonance field for $n = 2$ is also capable of exciting the $n = 6$ mode; thus, a secondary peak at $f_6$ = 17.44 GHz appears in the corresponding FFT spectrum in Fig. 7.

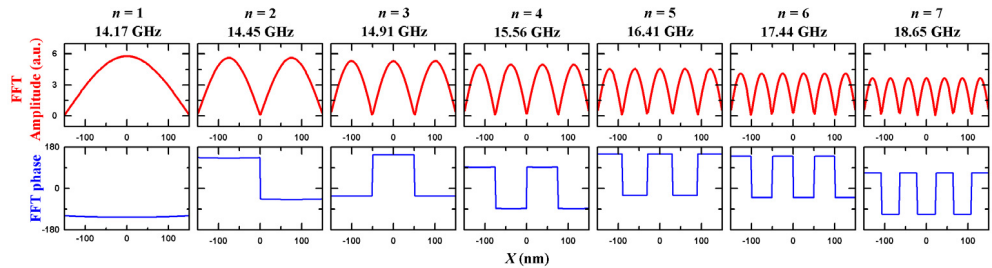

FIG. 8. FFT amplitude and phase diagrams for the seven normal modes of the magnetic nanowire. The modes show the typical standing wave nature of a one-dimensional oscillation system with a fixed boundary. For each mode, the nodes partition the space into zones of alternating phases, defining a unique space pattern.

Figure 8 presents the spatial distribution of the FFT amplitude and phase diagrams for the seven normal modes. We observe the clear standing wave nature of the modes, and the number of wave-function nodes corresponds to $n-1$. The abrupt phase change of $\pi$ across nodes divides the space into one-dimensional zones of alternating phases, making a space pattern that is unique to the index $n$. The width of these zones is defined by the distance between adjacent nodes, which is $300/n$ nm for corresponding mode index $n$, as expected by the fixed-boundary condition.

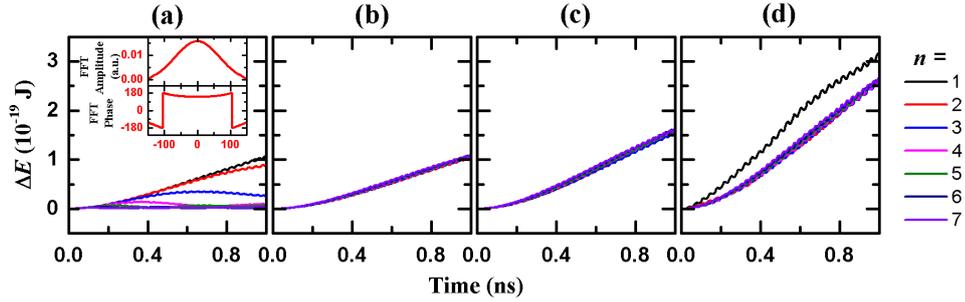

FIG. 9. Time evolution of the energy growth for the normal modes of the magnetic nanowire. The sample is in temporal resonance in (a). The inset shows the spatial distribution of the FFT amplitude and phase when the sample oscillation is in $n = 2$ temporal resonance. In (b), (c) and (d), the sample is in both temporal and spatial resonance. In addition, the external field fits the mode amplitude profile in (c) and is concentrated on the antinodes in (d).

Figure 9 displays the increase in total energy ($\Delta E$) of the seven normal modes under temporal and multidimensional resonance conditions. The oscillation amplitude of the external field is small (10 Oe) to suppress nonlinear dynamics. To make the results comparable, we keep the summation of the magnetic flux (absolute value) in all phase zones the same in the micromagnetic simulations. The temporal resonance in Fig. 9(a) shows the highest energy growth rate at $n = 1$ and a much lower rate as $n$ increases, with the exception of $n = 2$. The resonance frequency at $n = 2$ is very close to that at $n$

= 1. As a result, the spin wave excited by a globally uniform field at frequency $f_2$ resembles the fundamental mode rather than the $n = 2$ mode, as displayed in the inset of Fig. 9(a). However, we observe significant enhancement in growth of $\Delta E$ under multidimensional resonance for all modes except for the fundamental mode, as shown in Fig. 9(b). The spatial distribution of the external field matches the space pattern of the corresponding normal mode, but the field amplitude is uniform in each phase zone. The fundamental mode is the only mode that has a uniform phase distribution; therefore, multidimensional resonance and temporal resonance produce the same result at $n = 1$. The growth rate of $\Delta E$ increases by approximately 50% for all modes when the external field aligns with not only the space pattern but also the mode amplitude profile, as shown in Fig. 9(c). In Fig. 9(d), we observe the greatest energy growth rate when the external field is concentrated at the antinodes. The rate is approximately 100% higher for $n = 1$ and 67% higher for $n \geq 2$ than the results in Fig. 9(c).